\title{A resonance of the Higgs field at 700 GeV and a new phenomenology}
\author{
        \textsc{Maurizio Consoli} \\ INFN - Sezione di Catania,  I-95129 Catania, Italy \\
        {\tt{maurizio.consoli@ct.infn.it}}\\[0.5cm]
        \textsc{Leonardo Cosmai} \\ INFN - Sezione di Bari, I-70126 Bari, Italy \\
        {\tt{leonardo.cosmai@ba.infn.it}}
}

\documentclass[12pt]{article}

\usepackage[paper=a4paper,dvips,top=1.5cm,left=1.5cm,right=1.5cm,
    foot=1cm,bottom=1.5cm]{geometry}

\usepackage{times}

\usepackage[fleqn]{amsmath}
\usepackage{amsfonts}
\usepackage{amssymb}
\usepackage{amsthm}
\usepackage{amsopn}
\usepackage{xspace}
\usepackage{array}
\usepackage{epsfig}

\newcommand{\BE}{\begin{equation}}
\newcommand{\EE}{\end{equation}}

\begin{document}

\maketitle







%
%
%
%



\begin{abstract}
\noindent It has been recently proposed that, besides the known
resonance with mass $m_h\sim$ 125 GeV, the Higgs field could exhibit
a new excitation with a larger mass $M_h$ related by $M^2_h\sim
m^2_h \ln (\Lambda_s/M_h)$, where $\Lambda_s$ is the ultraviolet
cutoff of the scalar sector. Lattice simulations of the propagator
performed in the 4D Ising limit of the theory are consistent with
this two-mass picture and lead to the estimate $M_h\sim 700$ GeV. In
spite of its large mass, however, this heavier state would couple to
longitudinal vector bosons with the same typical strength of the
low-mass state and would thus represent a relatively narrow
resonance. In this Letter we argue that this hypothetical new
resonance would naturally fit with some excess of 4-lepton events
which is observed by ATLAS around 680 GeV.
\end{abstract}


\section{Introduction}

Spontaneous Symmetry Breaking (SSB) through the vacuum expectation
value $\langle \Phi \rangle \neq 0$ of a scalar field, is a basic
ingredient of the Standard Model. This old idea of a fundamental
scalar field \cite{Englert:1964et,Higgs:1964ia} has more recently
found an important experimental confirmation after the observation,
at the Large Hadron Collider of CERN
\cite{Aad:2012tfa,Chatrchyan:2012xdj} of a narrow scalar resonance,
of mass $m_h \sim 125 $ GeV whose phenomenology fits well with the
perturbative predictions of the theory. This has produced the
widespread conviction that, by now, any modification of this general
picture can only come from new physics.

Though, this is not necessarily true. So far only the gauge and
Yukawa couplings of the 125 GeV resonance have been tested. Instead,
the effects of a genuine scalar self-coupling $\lambda= 3
m^2_h/\langle \Phi \rangle^2$ are still below the accuracy of the
measurements. For this reason, an uncertainty about the mechanisms
producing SSB may still persist.

At the beginning, the driving mechanism was identified in a
classical scalar potential with a double-well shape. But, after
Coleman and Weinberg \cite{Coleman:1973jx}, the phenomenon started
to be described at the quantum level and the classical potential was
replaced by the effective potential $V_{\rm eff}(\varphi )$ which
includes the zero-point energy of all particles in the spectrum.

Yet, SSB could essentially be determined by the pure $\lambda\Phi^4$
sector if the other contributions to the vacuum energy are
negligible. As recently pointed out in ref.\cite{Consoli:2020nwb},
this becomes a natural perspective if one takes into account the
indications of most recent lattice simulations of pure
$\lambda\Phi^4$ in 4D
\cite{lundow2009critical,Lundow:2010en,akiyama2019phase}. These
calculations, performed in the Ising limit of the theory with
different algorithms, indicate that on the largest lattices
available so far the SSB phase transition is (weakly) first order,
as in the one-loop and Gaussian approximations to the effective
potential. In fact, differently from standard Renormalization-Group
improved perturbation theory, these approximations indicate that
massless $\lambda\Phi^4$ theory (i.e. classically scale invariant)
exhibits SSB and, therefore, the phase transition occurs earlier
when the quanta of the symmetric phase have a very small but still
positive mass squared $m^2_\Phi> 0$.

One then discovers that, in these approximations, $V_{\rm
eff}(\varphi )$ has {\it two} distinct mass scales:~i) a mass
$m^2_h$, defined by its quadratic shape at the minimum and~ ii) a
mass $M^2_h$, defined by the zero-point energy which determines its
depth. Always considered as being the same mass, in these
approximations one finds instead $M^2_h\sim L m^2_h \gg m^2_h$,
where $L=\ln (\Lambda_s/M_h)$ and $\Lambda_s$ is the ultraviolet
cutoff of the scalar sector. Since vacuum stability depends on the
much larger $M_h$, and not on $m_h$, SSB could originate within the
pure scalar sector regardless of the other parameters of the theory
(e.g. the vector boson and top quark mass).

Intuitively, it is clear that quadratic shape and depth of the
potential are different concepts. For a more formal explanation, we
just recall that the derivatives of the effective potential produce
(minus) the n-point functions at zero external momentum. Therefore
$m^2_h$, which is $V''_{\rm eff}(\varphi )$ at the minimum, is
directly the 2-point, self-energy function $|\Pi(p=0)|$. On the
other hand, the zero-point energy is (one-half of) the trace of the
logarithm of the inverse propagator $G^{-1}(p)=(p^2-\Pi(p))$.
Therefore, after subtracting constant terms and quadratic
divergences, matching the 1-loop zero-point energy(``$zpe$'') gives
the relation \BE\label{general} zpe\sim -\frac{1}{4} \int^{p_{\rm
max}}_{p_{\rm min}} {{ d^4 p}\over{(2\pi)^4}} \frac{\Pi^2(p)}{p^4}
\sim-\frac{ \langle \Pi^2(p)\rangle }{64\pi^2} \ln\frac{p^2_{\rm
max}}{p^2_{\rm min}}\sim -\frac{M^4_h}{64\pi^2} \ln\frac{\Lambda^2_s
}{M^2_h} \EE This shows that $M^2_h$ effectively includes the
contribution of all momenta and reflects a typical average value $
|\langle \Pi(p)\rangle| $ at larger $p^2$. In this sense, the
$m_h-M_h$ two-mass structure resembles the two branches (phonons and
rotons) in the energy spectrum of superfluid He-4 which is usually
considered the non-relativistic analog of the broken phase.

Note that the non uniform scaling of the two masses with $\Lambda_s$
is crucial not to run in contradiction with the ``triviality'' of
$\lambda\Phi^4$. This implies a continuum limit with a Gaussian set
of Green's functions and, thus, with just a massive free-field
propagator. With this constraint, a cutoff theory can either predict
$m_h\to M_h$ or a non-uniform scaling when $\Lambda_s \to \infty$.
The single-mass limit will then depend on the unit scale, $m_h$ or
$M_h$, chosen for measuring momenta \cite{Consoli:2020nwb}. Namely:
a) $m_h$ is the unit scale so that $M_h$ and the higher branch
simply decouple b) $M_h$ is the unit scale so that, when $\Lambda_s
\to \infty$, the phase space of the lower branch becomes smaller and
smaller until ideally shrinking to the zero-measure set $p_\mu=0$
which is transformed into itself under the Lorentz Group. This means
that the lower branch merges into the vacuum state and the only
remaining excitation is the higher branch with mass $M_h$.

The existence of a two-mass structure in the cutoff theory was
checked with lattice simulations of the scalar propagator
\cite{Consoli:2020nwb} in the Ising limit which is a convenient
laboratory to exploit the non-perturbative aspects of the theory. It
corresponds to a $\lambda\Phi^4$ with an infinite bare coupling
$\lambda_0=+\infty$, as if one were sitting precisely at the Landau
pole. For a given non-zero, low-energy coupling $\lambda \sim 1/L$,
this represents the best possible definition of the local limit with
a cutoff. Then, once $m^2_h$ is directly computed from the very
low-$p^2$ limit of $G(p)$ and $M^2_h$ is extracted from its
behaviour at higher $p^2$, the lattice data are consistent with a
transition between two different regimes and with the expected
increasing logarithmic trend $M^2_h\sim L m^2_h$.

If, for finite $\Lambda_s$, the scalar propagator really
interpolates between these two masses, by increasing the energy, one
could observe a transition from a relatively low value, e.g.
$m_h$=125 GeV, to a much larger $M_h$ \footnote{Two mass scales in
the propagator also require some interpolating form in loop
corrections. Since some precision measurements, e.g. $A_{FB}$ of the
b-quark or $\sin^2\theta_w$ from NC experiments
\cite{Chanowitz:2009dz}, still favor a rather large Higgs particle
mass, this could help to improve the overall quality of a Standard
Model fit.}. At the same time since, differently from $m_h$, the
larger mass $M_h$ would remain finite in units of the weak scale
$\langle \Phi \rangle\sim (G_{\rm Fermi}\sqrt{2})^{-1/2}\sim$ 246.2
GeV in the continuum limit, one can write a proportionality
relation, say $M_h=K \langle \Phi \rangle$, and extract the constant
$K$ from the same lattice data. As discussed in
\cite{Consoli:2020nwb}, this leads to a final estimate $M_h \sim 720
\pm 30 $ GeV which includes various statistical and theoretical
uncertainties and updates the previous work of
refs.\cite{Cea:2003gp, Cea:2009zc} \footnote{The value $M_h \sim 720
\pm 30 $ GeV includes the statistical errors in the $m_h/M_h$ ratio
from the lattice simulations of the propagator and the effect of
switching from leading-order to next-to-leading-order relations in
the $m_h-\langle \Phi\rangle$ interdependence. No uncertainty is
included from restricting the simulation to the Ising limit. Though,
within the simple 1-loop picture of a Landau pole, this remains the
best possible definition of the local theory at the lattice level.
Beyond 1-loop, standard perturbation theory gives contradictory
indications (Landau pole in odd orders vs. spurious ultraviolet
fixed points in even orders). Borel re-summation procedures
\cite{shirkov}-\cite{kazakov}, yielding a positive, monotonically
increasing $\beta-$function, support the idea of the Landau pole.}.

We emphasize that with the relation $M_h= K\langle \Phi \rangle$ we
are not introducing a new large coupling $3K^2=O(10)$ in the
phenomenology of SSB. This $3K^2$ is clearly quite distinct from the
other coupling $\lambda=3 m^2_h/ \langle \Phi \rangle^2\sim 1/L$ but
should not be viewed as a coupling constant which produces {\it
observable} interactions in the broken-symmetry phase. Since $M^4_h$
reflects the magnitude of the vacuum energy density, it would be
natural to consider $K^2\sim \lambda L$ as a {\it collective}
self-interaction of the vacuum condensate which persists in the
$\Lambda_s\to \infty$ limit. This original view
\cite{Consoli:1993jr,Consoli:1997ra} can intuitively be formulated
in terms of a scalar condensate whose increasing density $\sim L$
\cite{Consoli:1999ni} compensates for the decreasing strength
$\lambda \sim 1/L$ of the two-body coupling. On the other hand
$\lambda\sim 1/L$ remains as the appropriate coupling to describe
the {\it individual} interactions of the elementary excitations of
the vacuum, i.e. the Higgs field and the Goldstone bosons. In this
way, consistently with the ``triviality'' of $\lambda\Phi^4$ theory,
their interactions will become weaker and weaker for $\Lambda_s \to
\infty$.

With this description of the scalar sector, and by using the
Equivalence Theorem \cite{Cornwall:1974km,Chanowitz:1985hj}, the
same conclusion applies to the high-energy interactions of the Higgs
field with the longitudinal vector bosons in the full $g_{\rm
gauge}\neq 0$ theory. In fact, the limit of zero gauge coupling is
smooth \cite{Bagger:1989fc}. Therefore, up to corrections
proportional to $g_{\rm gauge}$, a heavy Higgs resonance will
interact exactly with the same strength as in the $g_{\rm gauge}= 0$
theory \cite{Castorina:2007ng}. For the convenience of the reader,
this point will be summarized in Sect.2. In Sect.3, we will instead
consider some phenomenological implications for the present LHC
experiments.

\section{Observable interactions for a large $M_h$}

Differently from here, where $m_h$ and $M_h$ are assumed to coexist,
in ref.\cite{Castorina:2007ng} it was adopted the asymptotic
single-mass limit b) of the Introduction which effectively
reproduces a standard propagator with mass $M_h$. In spite of this
difference, however, the problem was the same considered here: a
$\Lambda_s-$independent scaling law $M_h=K \langle \Phi \rangle$.
Then, is the constant $3K^2$ describing {\it observable}
interactions which survive in the $\Lambda_s \to \infty$ limit? If
so, what about ``triviality''?

As anticipated, the solution consists in interpreting $3K^2$ as a
collective self-coupling of the scalar condensate whose effects are
re-absorbed into the vacuum structure. As such, it is basically
different from the coupling $\lambda$ defined through the
$\beta-$function \BE \ln\frac{\mu}{\Lambda_s}=
\int^\lambda_{\lambda_0}~\frac{dx}{\beta(x)} \EE For  $\beta(x) =
3x^2/(16\pi^2) + O(x^3)$, whatever the bare contact coupling
$\lambda_0$ at the asymptotically large $\Lambda_s$, at finite
scales $\mu\sim M_h$ this gives $\lambda\sim 16\pi^2/(3L)$ with $L=
\ln (\Lambda_s/M_h)$. Therefore, consistently with ``triviality'',
this $\lambda$ (and not the $\Lambda_s-$ independent $3K^2$) is the
appropriate coupling to describe the interactions among the
fluctuations of the broken phase.

By introducing the W-mass $M_w=g_{\rm gauge}\langle \Phi \rangle/2 $
and with the notations of \cite{yndurain}, a convenient
parametrization \cite{Castorina:2007ng} of these residual
interactions in the scalar potential is ($r=M^2_h/4M^2_w =
K^2/g^2_{\rm gauge}$) \BE \label{potential} U_{\rm
scalar}=\frac{1}{2}M^2_h h^2 + \epsilon_1 r g_{\rm gauge} M_w
h(\chi^a\chi^a + h^2) +\frac{1}{8} \epsilon_2 r g^2_{\rm gauge}
(\chi^a\chi^a + h^2)^2 \EE The two parameters $\epsilon_1$ and
$\epsilon_2$ are usually set to unity and account for $\lambda \neq
3K^2$, i.e. \BE \epsilon^2_1=\epsilon_2= \frac{\lambda}{3K^2}\sim
1/L \EE Thus, one can consider that corner of the parameter space
\cite{Castorina:2007ng}, large $K^2$ but $M_h \ll \Lambda_s$, that
does not exist in the conventional view where one assumes $\lambda =
3K^2$.

But is this scenario still valid in the full gauge theory? In fact,
the original calculation \cite{lee} in the unitary gauge could give
the opposite impression, namely that, for a heavy $M_h$,
longitudinal $W_LW_L$ scattering is indeed governed by the large
coupling $ K^2 = M^2_h/\langle \Phi \rangle^2$. Since this is an
important point, we will repeat here the main argument of
\cite{Castorina:2007ng}.

In the unitary-gauge calculation of $W_LW_L \to W_LW_L$ high-energy
scattering, one starts from a tree-level amplitude $A_0$ which is
formally $O(g^2_{\rm gauge})$ but ends up with \BE A_0(W_LW_L \to
W_LW_L)\sim \frac{3 M^2_h g^2_{\rm gauge} }{4 M^2_w} = \frac{3M^2_h
}{\langle \Phi \rangle^2}=3K^2 \EE Here the factor $g^2_{\rm gauge}$
comes from the vertices. The $1/M^2_w$ derives from contracting with
the external longitudinal polarizations $\epsilon^{(L)}_\mu \sim
(k_\mu/M_w)$ and the factor $M^2_h$ emerges after expanding the
Higgs field propagator \BE \frac{1 }{s- M^2_h}\to \frac{1 }{s}(1 +
\frac{M^2_h }{s}+...) \EE While the leading $1/s$ contribution
cancels against a similar term from the other diagrams (which
otherwise would give an amplitude growing with $s$), the $M^2_h$
from the expansion of the propagator is effectively ``promoted'' to
the role of coupling constant. In this way, one gets exactly the
same result as in a pure $\lambda\Phi^4$ theory with a contact
coupling $\lambda_0=3K^2$.

However, this is only the tree approximation. To obtain the full
result, let us observe that the Equivalence Theorem is valid to all
orders in the scalar self-interactions \cite{Bagger:1989fc}.
Therefore we have not to worry to re-sum the troublesome infinite
vector-boson graphs. But, from the $\chi\chi \to \chi\chi$ amplitude
at a scale $\mu$ for $g_{\rm gauge}=0$ \BE A(\chi\chi \to \chi
\chi)\Big|_{g_{\rm gauge}=0}\sim \lambda \sim \frac{1 }{\ln
(\Lambda_s/\mu)}\EE we can deduce the result for the longitudinal
vector bosons in the ${g_{\rm gauge}\neq 0}$ theory, i.e.  \BE
A(W_LW_L \to W_LW_L)= [1 +O(g^2_{\rm gauge})]~A(\chi\chi \to \chi
\chi)\Big|_{g_{\rm gauge}=0}\sim \lambda\sim \frac{1 }{\ln
(\Lambda_s/\mu)}\EE This analysis, when extended to our present
perspective where $m_h$ and $M_h$ now coexist and could be
experimentally determined, indicates that at a scale $\mu \sim M_h$
the supposed strong interactions proportional to $\lambda_0= 3K^2$
are actually weak interactions controlled by the much smaller
coupling \BE \lambda= \frac{3m^2_h }{\langle \Phi \rangle^2 }= 3K^2
~\frac{m^2_h }{M^2_h}\sim 1/L \EE Analogously, the conventional very
large width into longitudinal vector bosons computed with the
coupling $\lambda_0= 3K^2$, say $\Gamma^{\rm conv}(M_h \to W_LW_L)
\sim M^3_h/\langle \Phi \rangle^2$, should instead be rescaled by
$\epsilon_1^2= (\lambda/3K^2)=m^2_h/M^2_h$. This gives \BE
\label{aux} \Gamma(M_h\to W_LW_L) \sim \frac{m^2_h }{M^2_h}
~\Gamma^{\rm conv}(M_h \to W_LW_L) \sim M_h ~\frac{m^2_h }{\langle
\Phi \rangle^2} \EE whose meaning is that $M_h$ indicates the
available phase space in the decay and ${m^2_h }/{\langle \Phi
\rangle^2} $ the interaction strength. Through the decays of the
heavier state, the scalar coupling $\lambda= 3m^2_h/\langle \Phi
\rangle^2$ could thus become visible. This would confirm that the
two masses $m_h$ and $M_h$ are not independent but represent
excitations of the same field.

\section{A new phenomenology}

After these preliminaries, which just summarize the results of
refs.\cite{Consoli:2020nwb} and \cite{Castorina:2007ng}, we arrive
to the main point of this Letter. Suppose to take seriously this
two-mass picture and the prediction of a second heavier excitation
of the Higgs field with mass $M_h\sim$ 700 GeV. Is there any
experimental indication for such a resonance? Furthermore, if there
were some potentially interesting signal, what kind of phenomenology
should we expect? Finally, is there any support for the
identification $m_h\sim$ 125 GeV, implicitly assumed for the
magnitude of our lower mass scale? In the following, we will
consider a definite piece of experimental evidence, namely a certain
excess of events observed by the ATLAS Collaboration
\cite{Aaboud:2017rel,ATLAS2} in the 4-lepton channel for an
invariant mass $\mu_{4l}\sim $ 700 GeV ($l=e,\mu$). In our
comparison, we will assume this region to be associated with our
heavier excitation of the Higgs field.

Of course, the 4-lepton channel is just one of the possible decay
channels of a hypothetical heavier Higgs resonance and, for a
comprehensive analysis, one should also look at the other final
states. For instance, at the 2-photon channel that, in the past, has
been showing some intriguing signal for the close energy of 750 GeV.
However, the 4-lepton channel, as compared to other final states,
has the advantage of being experimentally clean and, for this
reason, is considered the ``golden'' channel to explore the possible
existence of a heavy Higgs resonance. Moreover, the bulk of the
effect can be analyzed at an elementary level. Thus it makes sense
to start from here.

We emphasize that the main new aspect of our picture is the
substantial reduction of the conventional large width in
Eq.(\ref{aux}). Therefore, by using the conventional estimate for
$M_h=$ 700 GeV \cite{Djouadi:2005gi,handbook} \BE \label{th1}
\Gamma^{\rm conv}( M_h \to ZZ)\sim 56.7~{\rm GeV}\EE and from
Eq.(\ref{aux}) \BE \Gamma( M_h \to ZZ)\sim \frac{m^2_h }{M^2_h}
~\Gamma^{\rm conv}(M_h \to ZZ)\EE we could insert a given value for
$m_h$ and compute \BE \Gamma( M_h \to ZZ)\sim \frac{m^2_h}{(700~{\rm
GeV})^2}~56.7~{\rm GeV}\EE With the tentative identification $m_h=
125$ GeV, this gives \BE \Gamma( M_h \to ZZ)\sim 1.8~{\rm GeV} \EE
Afterward, to obtain the total width, we will maintain exactly the
other contributions reported in the literature
\cite{Djouadi:2005gi,handbook} for $M_h=$ 700 GeV, namely
\footnote{The quoted value refers to a top-quark mass of 173.7 which
averages between the two determinations 173.2(6) GeV and 174.2(1.4)
GeV reported by the Particle Data Group \cite{Tanabashi:2018oca} and
replaces the values 171.4 and 172.5 GeV used respectively in
\cite{Djouadi:2005gi} \cite{handbook}.} \BE\label{th3} \Gamma(M_h\to
{\rm fermions+ gluons+ photons...})\sim 28~{\rm GeV} \EE and the
same ratio \BE\label{th4} \frac{\Gamma( M_h \to WW)}{\Gamma( M_h \to
ZZ)} \sim 2.03 \EE These input numbers, which have very small
uncertainties, produce a total decay width \BE \label{th5}\Gamma(
M_h \to all) \sim 28~{\rm GeV} + 3.03~\Gamma (M_h \to ZZ) \sim
33.5~{\rm GeV} \EE and a branching fraction \BE B( M_h \to
ZZ)\sim\frac{\rm 1.8}{33.5}\sim 0.054 \EE Besides the decay modes of
the 700 GeV resonance, the other basic ingredient for its
phenomenology is the total inclusive cross section $\sigma (pp\to
M_h)$. The two main contributions derive from more elementary parton
processes where two gluons or two vector bosons $VV$  fuse to
produce the heavy state $M_h$ (here $VV=WW,ZZ$ would be emitted by
two quarks inside the protons). For this reason, the two types of
process are usually called Gluon-Gluon Fusion (GGF) and Vector-Boson
Fusion (VBF) mechanisms, i.e. \BE \sigma (pp\to M_h)\sim \sigma
(pp\to M_h)_{\rm GGF}+ \sigma (pp\to M_h)_{\rm VBF} \EE The
importance given traditionally to the latter process for large $M_h$
can be understood by observing that the $VV\to M_h$ process is the
inverse of the $M_h\to VV$ decay and therefore $\sigma(pp\to
M_h)_{\rm VBF}$ can be expressed \cite{kane} as a convolution with
the parton densities of the same Higgs resonance decay width.
Therefore, if its coupling to longitudinal $W$'s and $Z$'s were
proportional to $K^2=M^2_h / \langle \Phi \rangle^2$, with a
conventional width $\Gamma^{\rm conv}(M_h \to WW+ZZ) \sim$ 172 GeV
for $M_h\sim $ 700 GeV, the VBF mechanism would become sizeable
(about one-third of the GGF). But this coupling is not present in
our picture, where instead we expect \BE \Gamma(M_h \to WW+ZZ) \sim
\frac{m^2_h}{M^2_h}~ \Gamma^{\rm conv}(M_h \to WW+ZZ) \sim 5.5~{\rm
GeV}\EE As a consequence, the whole VBF will also be strongly
reduced from its conventional estimate $\sigma^{\rm conv}(pp\to
M_h)_{\rm VBF}=250\div300$ fb, i.e. \BE \sigma(pp\to M_h)_{\rm
VBF}\sim\frac{\rm 5.5}{172}~ \sigma^{\rm conv}(pp\to M_h)_{\rm
VBF}~\lesssim 10~ fb\EE This is well below the uncertainty of the
pure GGF contribution and will be neglected in the following.

Finally, the GGF term. Here, we will consider two slightly different
estimates. On the one hand, the value $\sigma (pp\to M_h)_{\rm
GGF}=800(80)$ fb of ref.\cite{Djouadi:2005gi} and on the other hand
the value $\sigma (pp\to M_h)_{\rm GGF}=1078(150)$ fb of
ref.\cite{handbook}. In both cases, errors include various
uncertainties due to the choice of the normalization scale and the
parametrization of the parton distributions. These predictions refer
to $\sqrt{s}=$ 14 TeV and should be rescaled by about $-12\%$ for
the actual center of mass energy of 13 TeV.

To compare with the data, we will approximate the cross section at
the resonance peak in terms of on-shell branching ratios as
\BE\label{exp3} \sigma_R (pp\to 4l)\sim \sigma (pp\to M_h)\cdot B(
M_h \to ZZ) \cdot 4 B^2( Z \to l^+l^-) \EE  since for a relatively
narrow resonance the effects of its virtuality should be small
\cite{goria}. In this relation, the $Z-$boson branching fraction
into charged leptons is precisely known and yields $4B^2( Z \to
l^+l^-)\sim 0.0045$. Altogether, for our reference value $B( M_h \to
ZZ)=$ 0.054, our predictions for the 4-lepton peak cross section and
the associated number of events (for luminosity of 36.1  and 139
fb$^{-1}$) are reported in Table 1.

\begin{table*}
\caption{\it For $M_h=$ 700 GeV and $m_h=$ 125 GeV, we report our
predictions for the peak cross section $\sigma_R(pp\to 4l)$ and the
number of events at two values of the luminosity. The two total
cross sections are our extrapolation to $\sqrt{s}=$ 13 TeV of the
values in \cite{Djouadi:2005gi} and \cite{handbook} for $\sqrt{s}=$
14 TeV. As explained in the text, only the GGF mechanism is relevant
in our model. }
\begin{center}
\begin{tabular}{cccc}
~~~~~~~~$\sigma(pp\to M_h)$ ~~~~~~~~~~~~&  $\sigma_R(pp\to 4l)$   ~~~~~~~& $n[4l]$ (36.1$fb^{-1})$ ~~~~&$n[4l]$(139$fb^{-1})$ \\
\hline
700(70)~fb   &  0.17(2)~fb  &$6.1\pm 0.6$  &  $23.6\pm 2.4$       \\
\hline
950(150)~fb   & 0.23(4)~fb  & $8.3\pm 1.3$ & $32.1 \pm 5.1$             \\
\hline
\end{tabular}
\end{center}
\end{table*}

For the smaller statistics of 36.1 fb$^{-1}$, see Fig.4a of
\cite{Aaboud:2017rel}, our results in Table 1 can be directly
compared, and are well consistent, with the measured value $ n_{\rm
4l}\sim (6\pm 3)$ for $\mu_{\rm 4l}=$ 700 GeV. Instead to compare
with the larger ATLAS sample \cite{ATLAS2} of 139 fb$^{-1}$, a
different treatment is needed. In fact, in their Fig.2d there are
now {\it three} points in the relevant energy region: at $\mu_{\rm
4l}\sim 635$ GeV, where $ n_{\rm 4l}\sim 7.0\pm 3.0 $, at $\mu_{\rm
4l}\sim 665$ GeV, where $ n_{\rm 4l}\sim 16.5\pm 4.0 $, and at
$\mu_{\rm 4l}\sim 695$ GeV, where $ n_{\rm 4l}\sim 9.0^{+2.2}_{-3.0}
$. Therefore, by defining $\mu_{4l}= E$ and $s=E^2$, we have assumed
that these 4-lepton events derive from the interference of a
resonating amplitude $A^{R}(s)\sim 1/(s - M^2_R)$ and a slowly
varying background $A^{B}(s)$. For positive interference below peak,
setting $M^2_R=M^2_h -i M_h \Gamma_h$, this gives a total cross
section \BE \label{sigmat} \sigma_T=\sigma_B
-\frac{2(s-M^2_h)~\Gamma_h M_h}{(s-M^2_h)^2+ (\Gamma_h
M_h)^2}~\sqrt{\sigma_B\sigma_R} +\frac{(\Gamma_h M_h)^2
}{(s-M^2_h)^2+ (\Gamma_h M_h)^2}~ \sigma_R\EE  where, in principle,
both the background cross section $\sigma_B$ and the resonating peak
$\sigma_R$ could be treated as free parameters. By converting the
ATLAS data into cross sections (for 139 fb$^{-1}$ and efficiency
$\sim 0.98$) and for 650 GeV $\leq M_h\leq $ 700 GeV, the best fit
is at $M_h=$ 682 GeV, see Fig.1. Since in our model, for small
changes of the mass, $\Gamma_h\equiv \Gamma(M_h \to all)$ varies
linearly with $M_h$, the width was correspondingly rescaled by the
relation $(\Gamma_h/M_h)=(33.5/700)$. ATLAS estimate of the
background is also shown as a dashed line. \vskip 30 pt
\begin{figure}[ht]
\begin{center}
\psfig{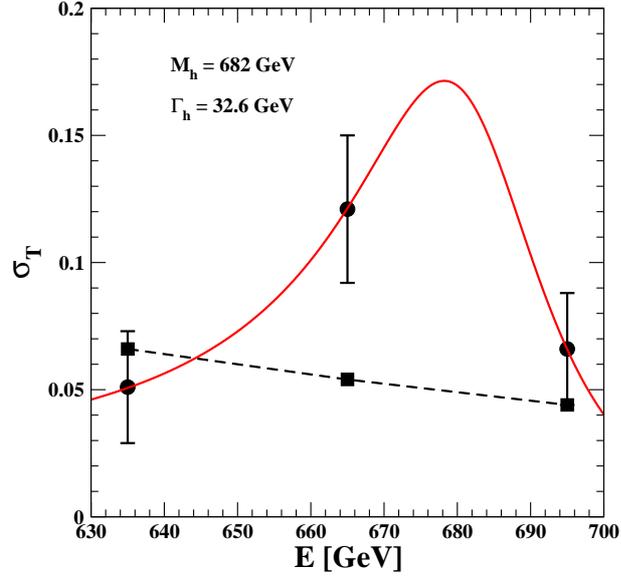}
\end{center}
\caption{ {\it For $M_h=$ 682 GeV and $\Gamma_h=$ 32.6 GeV we show a
fit (in red) with Eq.(\ref{sigmat}) to the ATLAS data converted into
cross sections in fb (the black dots). The fitted parameters are
$\sigma_B=0.009^{+0.017}_{-0.008}$ fb and $\sigma_R = 0.154\pm
0.054$ fb. Errors from fit correspond to $\Delta \chi^2=+$2.41 above
$\chi^2_{\rm min}=0$. The dashed line indicates ATLAS estimate of
the background.}}
\end{figure}

We observe that the best fit is found for very small background and
a peak cross section $\sigma_R$ consistent with our Table 1 where
$\sigma_R=$ 0.17(2) fb or $\sigma_R=$ 0.23(4) fb. Therefore, from
this simple exercise, we deduce the following conclusions:

1) a fit with just the ATLAS estimate of the background has much
larger value $\chi^2=6.5$ and indicates the presence of some
additional contribution

2) when looking for this additional contribution as a new resonance,
and leaving the background as free parameter, a fit to the data
provides a pair of values $(M_h,\Gamma_h)$ and a resonance peak
which are consistent with our predictions. In particular, $M_h$
tends to stay in the lower theoretical band $M_h=690\pm 10 ~({\rm
stat}) \pm 20 ~({\rm sys})$ GeV obtained when our lattice data
\cite{Consoli:2020nwb} are combined with the leading-order estimate
for the $m_h-\langle\Phi\rangle$ interdependence \footnote{Setting
$M^2_h= m^2_h L \cdot(c_2)^{-1}$ and $m^2_h L =(16\pi^2/9) \cdot
\langle \Phi \rangle^2$ gives $M_h= (4\pi/3)\cdot (c_2)^{-1/2}
\langle \Phi \rangle$. From our lattice data \cite{Consoli:2020nwb}
we found a constant $(c_2)^{-1/2} = 0.67 \pm 0.01 ~({\rm stat}) \pm
0.02 ~({\rm sys})$.}.

Therefore, given the special role of the 4-lepton channel, further
checks of the background and further statistical tests would be
desirable. For instance, with the statistics of 36.1 fb$^{-1}$, the
deep diving of the local $p_0$ \cite{denys}, at the 3.8 sigma level,
was already supporting the existence of a new narrow resonance
around 700 GeV. It remains to be seen if an unambiguous answer could
still be obtained with the present LHC configuration or has to be
postponed to the future high-luminosity phase.



\end{document}